\documentclass[twocolumn,showpacs,preprintnumbers]{revtex4}
\usepackage{amssymb}
\usepackage{graphicx}
\usepackage{dcolumn}
\usepackage{bm}

\begin{document}

\title{Quantum dynamics of bound entangled states}
\author{Guo-Qiang Zhu$^{1}$ and Xiaoguang Wang$^{1,2}$}
\affiliation{1,Zhejiang Institute of Modern Physics, Zhejiang University, Hangzhou, P.R.
China}
\affiliation{2, Department of Physics and Institute of Theoretical Physics, The Chinese
University of Hong Kong, Hong Kong, China}
\date{\today }

\begin{abstract}
By using the partial transpose and realignment method,we study the time
evolution of the bound entanglement under the bilinear-biquadratic
Hamiltonian. For the initial Horodecki's bound entangled state, it keeps
bound entangled for some time, while for the initial bound entangled states
constructed from the unextendable product basis, they become free once the
time evolution begins. The time evolution provides a new way to construct
bound entangled states, and also gives a method to free bound entanglement.
\end{abstract}

\maketitle





Quantum information theory has drawn much attention in the last decade, and
in the quantum information theory, quantum entanglement plays an important
role as it can be utilized to realize some quantum processes \cite{nielsen}.
In realistic world, due to the interaction between the system and the
environment, most states in nature are mixed and the set of mixed states is
dense in Hilbert space. The mixed entangled states can be divided into two
classes \cite{horodecki}, one is free, which means that the state can be
distilled, the other is bound, which means that the state cannot be
distilled. The bound entangled state is thought to be useless in quantum
communication. However, the bound entangled state (BES) can produce a
nonclassical effect, which is called activation of bound entanglement. The
underlying concept originates from a formal entanglement-energy analogy \cite%
{en-en,acta,rohrlich}. It implies that the bound entanglement is like the
energy of a system confined in a shallow potential well. If we add a small
amount of extra energy to the system, its energy can be deliberated. One of
the main consequence of the existence of BESs is that it reveals a
transparent form of irreversibility in entanglement processing \cite{review}%
. The irreversibility can be viewed as an analog to irreversible
thermodynamics processes \cite{thermal}.

The bound entanglement can possibly get free if we add some energy
to it. In this paper, we see what happens if we let the BES
undergoes some quantum dynamical evolution under physical
Hamiltonians. We will see that BESs can be free via time evolution.
The time-evolved BES can be also considered as the original BES with
an extra time parameter $t$, and this provides a way to construct
more general BESs.

There are some measures to detect entangled state in
high-dimensional Hilbert space. For example, Peres-Horodecki
criterion based on partial transpose (PT)
\cite{PPT1,PPT2,negativity} and the realignment criterion
\cite{realign1,realign2}. Consider a density matrix $\rho $ in a
$n\otimes n$ system $,$ the PT with respect to the second system,
and the realignment of
the density matrix is given by $(\rho ^{T_{2}})_{ij,kl}=\rho _{il,kj}$ $%
(\rho ^{R})_{ij,kl}=\rho _{ik,jl}$ , respectively. Two quantities are
defined as
\begin{equation}
N_{1}=\frac{\parallel \rho ^{T_{2}}\parallel -1}{2},\;N_{2}=\frac{\parallel
\rho ^{R}\parallel -1}{2},  \label{negativity}
\end{equation}%
where the first is the negativity \cite{negativity}. The trace norm $%
\parallel A\parallel $ is given by $\parallel A\parallel =$tr$\sqrt{%
AA^{\dagger }}$. Either $N_{1}>0$ or $N_{2}>0$ indicates that the state is
entangled, $N_{1}=0$ and $N_{2}>0$ indicates that the state is bound
entangled, and $N_{1}>0$ means that the state is free entangled~\cite{Simon}.

To study time evolution, \ we should have a suitable Hamiltonian. We will
concentrate on $3\otimes 3$ system in this study, and without loss of
generality, regard this system as a two spin-one system. The most
representative interacting model in spin-one system is the
bilinear-biquadratic model~\cite{affleck}. The corresponding Hamiltonian is
given by~\cite{affleck}.
\begin{equation}
H=\mathbf{s}_{1}\cdot \mathbf{s}_{2}-\beta (\mathbf{s}_{1}\cdot \mathbf{s}%
_{2})^{2}.  \label{BB}
\end{equation}%
Here, $\mathbf{s}_{i}\ $denotes the spin-1 operator. As we know, for spin-1
system, the swapping operator $\mathbf{S}_{12}$ can be written as \cite%
{xgwang}
\begin{equation}
\mathbf{S}_{12}=(\mathbf{s}_{1}\cdot \mathbf{s}_{2})^{2}+\mathbf{s}_{1}\cdot
\mathbf{s}_{2}-1,
\end{equation}%
which is invariant under the $SU(3)$ unitary transformation. The singlet
projection operator is written as
\begin{equation}
\mathbf{P}_{12}=\frac{1}{3}\left[ (\mathbf{s}_{1}\cdot \mathbf{s}_{2})^{2}-1%
\right] .
\end{equation}%
Then, the Hamiltonian can be rewritten in terms of swapping operator and
projective operator,
\begin{equation}
H=\mathbf{S}_{12}-3(1+\beta )\mathbf{P}_{12}-\beta.
\end{equation}

It is obvious that swapping operator and projective operator are
commutative, so that the unitary matrix $U(t)$ can be obtained as
\begin{equation}
U(t)=e^{-iHt}=e^{-it\mathbf{S}_{12}}e^{i3(1+\beta )t\mathbf{P}%
_{12}}e^{it\beta}.
\end{equation}%
We have $\mathbf{S}_{12}^{2}=1$ and $\mathbf{P}_{12}^{2}=\mathbf{P}_{12}$,
then
\begin{eqnarray}
e^{-it\mathbf{S}_{12}} &=&\cos t-i\sin t\mathbf{S}_{12},  \nonumber \\
e^{i3(1+\beta )t\mathbf{P}_{12}} &=&1+(e^{i3(1+\beta )t}-1)\mathbf{P}_{12}.
\end{eqnarray}%
In this way, we can get the evolution operator exactly.

For clarity, we consider the case of $\beta =-1$. The Hamiltonian can be
reduced to the swap
\begin{equation}
H=\mathbf{S}=\mathbf{S}_{12}
\end{equation}%
up to a additive constant. In the following, we will study the
time-evolution of bound entangled state govenered by the Hamiltonian. The
initial state will evolve under the unitary density matrix
\begin{equation}
U(t)=e^{-itH}=\cos t-i\sin t\mathbf{S}.  \label{unitary}
\end{equation}%
At time $t$, the density matrix $\rho (t)=e^{-itH}\rho (0)e^{itH}$.

The two-spin BES we are considering, described by Horodecki in Ref.~\cite%
{horodecki}, is given by
\begin{equation}
\rho _{\alpha }(0)=\frac{2}{7}\mathbf{P}_{+}+\frac{\alpha }{7}\varrho _{+}+%
\frac{5-\alpha }{7}\varrho _{-},\ 2\leq \alpha \leq 5,  \label{initial}
\end{equation}%
where
\begin{eqnarray}
\mathbf{P}_{+} &=&|\Psi _{+}\rangle \langle \Psi _{+}|,|\Psi _{+}\rangle =%
\frac{1}{\sqrt{3}}(|00\rangle +|11\rangle +|22\rangle ),  \nonumber \\
\varrho _{+} &=&\frac{1}{3}(|01\rangle \langle 01|+|12\rangle \langle
12|+|20\rangle \langle 20|),  \nonumber \\
\varrho _{-} &=&\frac{1}{3}(|10\rangle \langle 10|+|21\rangle \langle
21|+|02\rangle \langle 02|)=\mathbf{S}\varrho _{+}\mathbf{S}.  \nonumber
\end{eqnarray}%
In Ref. \cite{horodecki}, Horodecki have demonstrated that

\[
\rho _{\alpha }\text{ is }\left\{
\begin{array}{l}
\text{separable for }2\leq \alpha \leq 3 \\
\text{bound entangled for }3<\alpha \leq 4, \\
\text{free entangled for }\alpha <4\leq 5.%
\end{array}%
\right.
\]%
An interesting feature of this state is that it is invariant under the joint
transformation of exchange $\alpha \leftrightarrow 5-\alpha $ and swap of
two particles. It is clear that the density matrix $\rho _{\alpha }$ has
four distinct eigenvalues, which are $2/7$, $0$, $(5-\alpha )/21$, $\alpha
/21$. They are non-degenerate, 2-fold degenerate, 3-fold degenerate, and
3-fold degenerate, respectively.

{}From the expression of the state, after PT, the trace norm of the
partially transposed state is given by
\begin{eqnarray*}
\left\Vert \rho _{\alpha }^{T_{2}}\right\Vert &=&\frac{1}{7}\Big[2+\frac{1}{2%
}(5+\sqrt{41-4\alpha (5-\alpha )} \\
&&+\left\vert 5-\sqrt{41-4\alpha (5-\alpha )}\right\vert )\Big]\text{ } \\
&=&\text{ }\left\{
\begin{array}{l}
1\text{ for }2\leq \alpha \leq 4, \\
\frac{1}{7}\left( 2+\sqrt{41-4\alpha (5-\alpha )}\right) \text{ for }\alpha
<4\leq 5.%
\end{array}%
\right.
\end{eqnarray*}%
Note that the realignment in the realignment criteria is equivalent to a
one-side swap followed by a PT \cite{Fan}, namely, $\left\Vert \rho _{\alpha
}^{R}\right\Vert =\left\Vert \left( S\rho _{\alpha }\right)
^{T_{2}}\right\Vert .$ It is direct to show that
\[
S\rho _{\alpha }=\frac{2}{7}\mathbf{P}_{+}+\frac{\alpha }{7}\mathbf{S}%
\varrho _{+}+\frac{5-\alpha }{7}\varrho _{+}\mathbf{S}.\
\]%
Then, after PT, we have

\[
\left\Vert \rho _{\alpha }^{R}\right\Vert =\frac{1}{21}\left( 19+2\sqrt{%
19-3\alpha (5-\alpha )}\right) .
\]%
From the expression of the two trace norms, an observation is that the
exchange $\alpha \longleftrightarrow (5-\alpha )$ does not change the values
of the two trace norms.

Now we consider the time evolution of state $\rho _{\alpha }.$ From the
evolution operator (\ref{unitary}) and the initial state (\ref{initial}),
the state at time $t$ is obtained as
\begin{eqnarray}
\rho _{\alpha }(t) &=&\frac{1}{7}\{2\mathbf{P}_{+}+a(t)\varrho
_{+}+b(t)\varrho _{-}+ic(t)[\mathbf{S},\varrho _{-}]\}, \\
a(t) &=&\alpha \cos ^{2}t+(5-\alpha )\sin ^{2}t  \nonumber \\
b(t) &=&\alpha \sin ^{2}t+(5-\alpha )\cos ^{2}t\   \nonumber \\
c(t) &=&\sin t\cos t(2\alpha -5)=1/2\sin (2t)(2\alpha -5)  \label{abc1}
\end{eqnarray}%
In the derivation of the above equation, we have used the identity $[\mathbf{%
S},\varrho _{-}]=[\varrho _{+},\mathbf{S}].$ It is obvious that
\begin{equation}
a+b=5,a-b=(2\alpha -5)\cos (2t),  \label{abc2}
\end{equation}
which will be used later.

Let's study the case in which $t=\pi /2$. At this moment,
\begin{equation}
U(t=\pi /2)=-i\mathbf{S}.
\end{equation}%
So the density matrix becomes
\begin{eqnarray}
\rho _{\alpha }(\pi /2) &=&U(\pi /2)\rho _{\alpha }U^{\dagger }(\pi /2)
\nonumber \\
&=&\frac{2}{7}\mathbf{P}_{+}+\frac{\alpha }{7}\varrho _{-}+\frac{5-\alpha }{7%
}\varrho _{+}.\
\end{eqnarray}%
The above state can be obtained from $\rho _{\alpha }(0)$ by just exchange $%
\alpha \longleftrightarrow (5-\alpha ).$ But the exchange does not affect
the values of trace norms. Thus, the trace norms at time $t=\pi /2$ are the
same as those at time $t=0$.

The negativity can be calculated by first making the PT, and diagonalizing
the partially transposed density matrix. Making use of the following
properties,

\begin{eqnarray}
\mathbf{P}_{+}^{T_{2}} &=&\frac{\mathbf{S}}{3},\varrho _{\pm
}^{T_{2}}=\varrho _{\pm },  \nonumber \\
(\mathbf{S}\rho _{-})^{T_{2}} &=&\frac{1}{3}\left( |00\rangle \langle
11|+|11\rangle \langle 22|+|22\rangle \langle 00|\right)  \nonumber \\
(\rho _{-}S)^{T_{2}} &=&\frac{1}{3}\left( |11\rangle \langle 00|+|22\rangle
\langle 11|+|00\rangle \langle 22|\right) ,  \label{ptt}
\end{eqnarray}%
we have

\begin{eqnarray}
\rho _{\alpha }^{T_{2}}(t) &=&\frac{1}{21}\{2\mathbf{S}+3a(t)\varrho
_{+}+3b(t)\varrho _{-}  \nonumber \\
&&+ic(t)\left( |00\rangle \langle 11|+|11\rangle \langle 22|+|22\rangle
\langle 00|\right) \\
&&-ic(t)\left( |11\rangle \langle 00|+|22\rangle \langle 11|+|00\rangle
\langle 22|\right) \}.\ \
\end{eqnarray}

In the basis spanned by $\{|00\rangle ,|11\rangle ,|22\rangle ,|01\rangle
,|10\rangle ,|12\rangle ,|21\rangle ,|20\rangle ,|02\rangle \},$ matrix $%
\rho _{\alpha }^{T_{2}}(t)$ can be written in a block-diagonal form

\begin{eqnarray*}
&&\rho _{\alpha }^{T_{2}}(t) =\frac{1}{21}\left( A_{_{3\times 3}}\oplus
B_{2\times 2}\oplus B_{2\times 2}\oplus B_{2\times 2}\right) , \\
&&A_{_{3\times 3}} =\left(
\begin{array}{ccc}
2 & ic(t) & -ic(t) \\
-ic(t) & 2 & ic(t) \\
ic(t) & -ic(t) & 2%
\end{array}%
\right) ,B_{2\times 2}=\left(
\begin{array}{cc}
a(t) & 2 \\
2 & b(t)%
\end{array}%
\right) ,
\end{eqnarray*}%
where we have used the fact that in this basis, the swap can be written as

\[
\mathbf{S}=I_{3\times 3}\oplus \sigma _{x}\oplus \sigma _{x}\oplus \sigma
_{x}.
\]%
and $I_{_{3\times 3}}$ is a 3$\times 3$ identity matrix, and $\sigma _{x}$
is the $x$ component of the Pauli matrix vector.

Having the block-diagonal form of $\rho ^{T_{2}}(t),$one can obtain the
eigenvalues of the partial transposed density matrix, and only the two
following eigenvalues are possibly negative
\begin{eqnarray}
\lambda _{1} &=&\frac{1}{21}\left( 2-\sqrt{3}c\right)  \nonumber \\
&=&\frac{1}{42}\left( 4-\sqrt{3}|\sin (2t)(2\alpha -5)|\right) .  \nonumber
\\
\lambda _{2} &=&\frac{1}{42}(a(t)+b(t)-\sqrt{[a(t)-b(t)]^{2}+16})  \nonumber
\\
&=&\frac{1}{42}\left( 5-\sqrt{16+(2\alpha -5)^{2}\cos ^{2}2t}\right) ,
\label{neg1}
\end{eqnarray}%
So, the negativity of $\rho _{\alpha }(t)$ is given by
\begin{equation}
N_{1}=\max (0,-\lambda _{1})+3\max (0,-\lambda _{2}).  \label{neg2}
\end{equation}

Having known analytical expression of $N_{1},$ we next consider the quantity
$N_{2}.$ First, we need to obtain matrix $(S\rho _{\alpha }(t))^{T_{2}}.$
From the definition of the BES, we have
\[
\mathbf{S}\rho _{\alpha }(t)=\frac{1}{7}\{2\mathbf{P}_{+}+a(t)\mathbf{S}%
\varrho _{+}+b(t)\mathbf{S}\varrho _{-}+ic(t)\left( \varrho _{-}-\varrho
_{+}\right) \}.\
\]%
After PT, we obtain%
\begin{eqnarray*}
(S\rho _{\alpha }(t))^{T_{2}} &=&\frac{1}{21}\{2\mathbf{S}+i3c(t)\left(
\varrho _{-}-\varrho _{+}\right) \\
&&+a(t)\left( |11\rangle \langle 00|+|22\rangle \langle 11|+|00\rangle
\langle 22|\right) \\
&&+b(t)\left( |00\rangle \langle 11|+|11\rangle \langle 22|+|22\rangle
\langle 00|\right) \},\
\end{eqnarray*}%
where Eq. (\ref{ptt}) is used. In the basis given above, the matrix $(S\rho
_{\alpha }(t))^{T_{2}}$ can be written as
\begin{eqnarray*}
&&(S\rho _{\alpha }(t))^{T_{2}} =\frac{1}{21}\left( C_{_{3\times 3}}\oplus
D_{2\times 2}\oplus D_{2\times 2}\oplus D_{2\times 2}\right) , \\
&&C_{_{3\times 3}} =\left(
\begin{array}{ccc}
2 & b(t) & a(t) \\
a(t) & 2 & b(t) \\
b(t) & a(t) & 2%
\end{array}%
\right) ,D_{2\times 2}=\left(
\begin{array}{cc}
-ic(t) & 2 \\
2 & ic(t)%
\end{array}%
\right)
\end{eqnarray*}%
Again, it is of block-diagonal form. The square roots of eigenvalues of $%
CC^{\dagger }$ are given by
\begin{eqnarray}
\xi _{1} &=&2+a+b=7,  \nonumber \\
\xi _{2} &=&\xi _{3}=\sqrt{\left( 2-\frac{a+b}{2}\right) ^{2}+\frac{3}{4}%
(a-b)^{2}}  \nonumber \\
&=&\frac{1}{2}\sqrt{1+3(2\alpha -5)^{2}\cos ^{2}(2t)},
\end{eqnarray}%
and the square roots of eigenvalues of $DD^{\dagger }$ are
\begin{eqnarray}
\zeta _{1} &=&|2+c|=\frac{1}{2}|4+\sin (2t)(2\alpha -5)|,  \nonumber \\
\zeta _{2} &=&|2-c|=\frac{1}{2}|4-\sin (2t)(2\alpha -5)|.
\end{eqnarray}%
Thus, one has the trace norm
\begin{eqnarray}
\left\Vert \rho ^{R}\right\Vert &=&\frac{1}{3}+\frac{1}{21}\sqrt{1+3(2\alpha
-5)^{2}\cos ^{2}(2t)}  \nonumber \\
&&+\frac{1}{14}(|4+\sin (2t)(2\alpha -5)|  \nonumber \\
&&+|4-\sin (2t)(2\alpha -5)|).  \label{normm}
\end{eqnarray}%
From the expression of the trace norm, when $2\leq \alpha \leq 4.5,$ the
above expression can be simplified as

\begin{equation}
\left\Vert \rho ^{R}\right\Vert =\frac{1}{21}\left( 19+\sqrt{1+3(2\alpha
-5)^{2}\cos ^{2}(2t)}\right)
\end{equation}%
for any time. Quantity $N_{2}$ is obtained by substituting Eq. (\ref{normm})
to (\ref{negativity}).

\begin{figure}[tbp]
\includegraphics[width=0.45\textwidth]{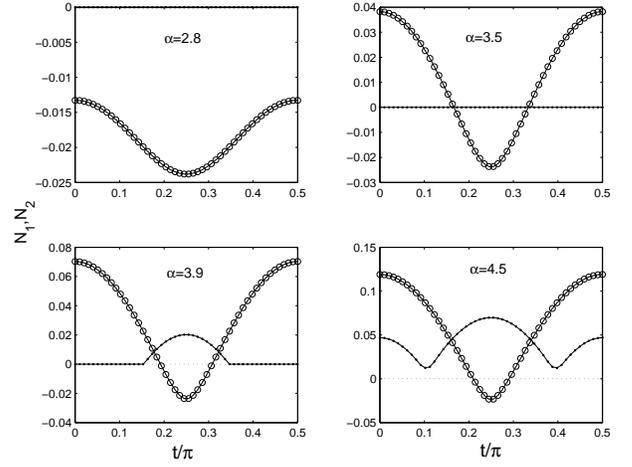}
\caption{Time evolution of the negativity $N_1$ and the quantity $N_2$
versus $t$ with initial state being Horodecki's state.}
\end{figure}

Now, we analyze the analytical results of negativity $N_{1}$ and quantity $%
N_{2}.$ One can see that when $t=\pi /2$, $N_{1}$ and $N_{2}$ are both back
to the value when $t=0$, because the swap operator has no effect on the
entanglement of a bipartite system. When $t=\pi $, the state has undergone a
cycle. One can find when $\alpha =5/2$, all the $\lambda _{i}$ keep constant
with the time. This is due to the fact that the BES with $\alpha =5/2$ is
invariant under swapping operation.

From Eqs. (\ref{neg1}) and (\ref{neg2}), it is not difficult to find that,
for arbitrary time, $N_{1}=0$ when $2\leq \alpha \leq \frac{5}{2}+\frac{2}{%
\sqrt{3}}\approx 3.655.$ Then, we know that in this range if $N_{2}>0$, the
state is bound entangled. In Fig.~1, we plot the negativity $N_{1}$ and the
quantity $N_{2}$ versus $t$ for different $\alpha $. When the initial state
is a separable state ($\alpha =2.8$), one cannot generate entanglement. For $%
\alpha =3.5$, the initial state is bound entangled, and as time increases,
the strength of bound entanglement decreases until it vanishes. After some
time, the state becomes bound entangled again. Here, there is no free
entanglement.

When we increase $\alpha$ to $\alpha=3.9>3.655$, the state can be made free
entangled over a certain range of time within one period. After time
evolution begins, The entanglement keep bounded until it becomes free. From
the figure, we also see that when the initial state is free entangled ($%
\alpha=4.5$), the entanglement keeps free all the time. We see that the
bound entanglement can be free by the time evolution.

Now, let us see the evolution of concurrence. We cannot get concurrence, but
can know a lower bound. In Ref.~\cite{kchen}, Chen et al. derived a theorem
that for any $n\otimes n$ mixed quantum state $\rho $, the concurrence $%
C(\rho )$satisfies
\begin{equation}
C(\rho )\geq \sqrt{\frac{2}{n(n-1)}}[\max (\parallel \rho ^{T_{2}}\parallel
,\parallel R(\rho )\parallel )-1],
\end{equation}%
The theorem gives the lower bound of concurrence of an arbitrary mixed
state. In the case of $n=3$. From Fig.~1, we can read the time behaviors of
the lower bound. For $\alpha =3.9,4.5$, the bound is always larger than
zero, and it displays some singularities due to the competition between the
two trace norms.

\begin{figure}[tbp]
\includegraphics[width=0.45\textwidth]{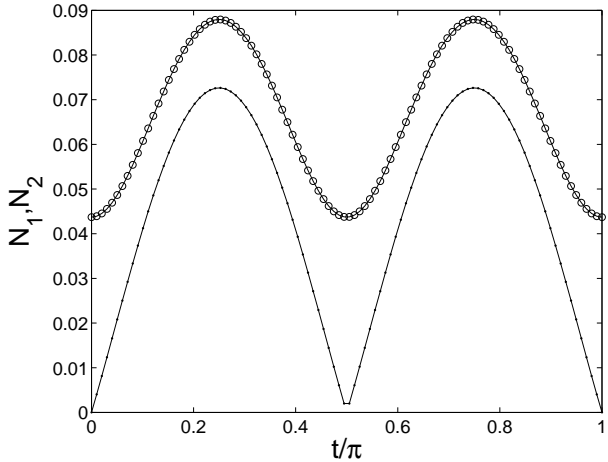}
\caption{Time evolution of the negativity $N_1$ and the quantity $N_2$
versus $t$ with initial state being first BES constructed from UPB.}
\end{figure}

Next, we consider BES constructed from the unextendable product basis (UPB).
The first BES from UPB is given by~\cite{bennet},
\begin{eqnarray}
|\phi _{0}\rangle &=&\frac{1}{\sqrt{2}}|0\rangle (|0\rangle -|1\rangle ),\ \
|\phi _{1}\rangle =\frac{1}{\sqrt{2}}(|0\rangle -|1\rangle )|2\rangle ,
\nonumber \\
|\phi _{2}\rangle &=&\frac{1}{\sqrt{2}}|2\rangle (|1\rangle -|2\rangle ),\ \
|\phi _{3}\rangle =\frac{1}{\sqrt{2}}(|1\rangle )-|2\rangle )|0\rangle ,
\nonumber \\
|\phi _{4}\rangle &=&\frac{1}{3}(|0\rangle +|1\rangle +|2\rangle )(|0\rangle
+|1\rangle +|2\rangle ),
\end{eqnarray}%
from which the density matrix could be expressed as

\begin{equation}
\rho _{_{UPB}}=\frac{1}{4}(I_{9\times 9}-\sum_{j=0}^{4}|\phi _{j}\rangle
\langle \phi _{j}|),  \label{upb}
\end{equation}%
here $I_{9\times 9}$ is the $9\times 9$ identity matrix. The second BES from
UPB is given by~Eq. (\ref{upb}) with \cite{bennet}%
\begin{eqnarray*}
|\phi _{j}\rangle &=&|\vec{v}_{j}\rangle \otimes |\vec{v}_{2j\text{mod}%
5}\rangle ,j=0,...4, \\
|\vec{v}_{j}\rangle &=&\frac{2}{\sqrt{5+\sqrt{5}}}\left[ \cos (2\pi
j/5),\sin (2\pi j/5),\sqrt{1+\sqrt{5}}/2\right] .
\end{eqnarray*}%
The negativity of both these two states are zero, but the trace norm is
given by 1.087 and 1.098, respectively, indicating that these two states are
bound entangled.

\begin{figure}[tbp]
\includegraphics[width=0.45\textwidth]{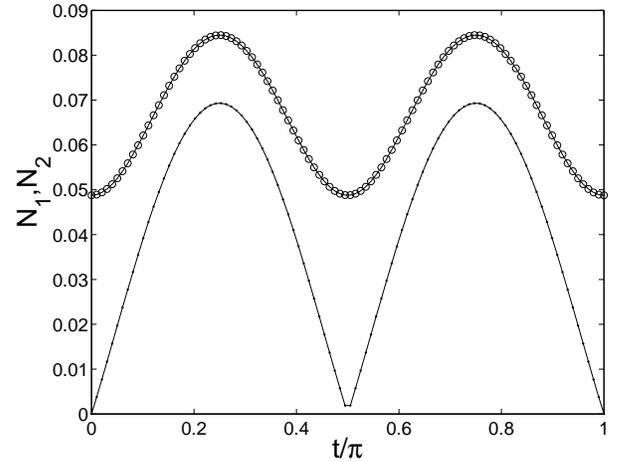}
\caption{Time evolution of the negativity $N_1$ and the quantity $N_2$
versus $t$ with initial state being second BES constructed from UPB.}
\end{figure}

The time evolution of the two BESs are plotted in Fig.~2 and 3,
respectively. There is no qualitative difference between the two plots. The
bound entanglement becomes free once triggering the time evolution, namely,
the bound entanglement is very fragile comparing with that in the
Horodecki's state. Also, we see that the lower bound of concurrence is
determined by quantity $N_{2}$ for all time. We can free bound entanglement
by the time evolution.


In conclusion, we have studied the time-evolution of BESs by two different
operational approaches, namely, the partial transpose and realignment
approach. The entanglement properties of time-evolution of different kind of
initial BESs under the bilinear-biquadratic Hamiltonian. The analytical
results of trace norms are obtained for the initial state being Horodecki's
BES at any time.

For the initial Horodecki's BES, the state keeps bound entangled in the
beginning of time evolution, while for the initial BESs constructed from
UPB, the BES becomes free once the time evolution begins. The time behaviors
are very different for different `class' of BESs. Some BESs are relatively
stable against time evolution, and some are fragile. Even a small
perturbation can free the bound entanglement from UPB. The time evolution
provide a new way to construct BES, and also gives a method to free bound
entanglement.

This work was supported by NSFC No. 10674117 and 10405019, specialized
Research Fund for the Doctoral Program of Higher Education (SRFDP) under
grant No.20050335087, and the project-sponsored by SRF for ROCS and SEM.
This work is partially supported by the Direct Grant of CUHK (A/C 2060286).%
%

\end{document}